\newcommand{\beq}{\begin{quote}}
\newcommand{\enq}{\end{quote}}
\newcommand{\be}{\begin{equation}}
\newcommand{\en}{\end{equation}}

\documentstyle[12pt]{article}
\begin{document}
\title{ Critique of
{\it ``Quantum Enigma: Physics encounters Consciousness''}} 

\date{}
\author{Michael Nauenberg\\
Department of Physics\\
University of California, Santa Cruz, CA 95064 
}
\maketitle

\begin{abstract}
The  central claim 
that understanding quantum mechanics
requires a conscious observer, which is made
made by B. Rosenblum and F. Kuttner
in their book "Quantum Enigma: Physics encounters
consciousnes", is shown to be based on various 
misunderstandings and distortions of the foundations of quantum mechanics.
\end{abstract}
\subsection*{Introduction}

When discussing  the  quantum theory
and its interpretation in  physics, Bohr often emphazised the importance
of describing  fully the experimental apparatus. Part of such a description
consists of selecting the proper 
words to describe the observations. In 1948 he put it as follows \cite{bohr2}:
\beq 
{\sl Phrases often found in the physical literature as `disturbance of phenomena
by observation' or `creation of physical attributes of objects by measurements'
represent a use of words like `phenomena' and `observation' as well as
`attribute' and `measurement' which is hardly compatible with common usage and
practical definition and, therefore, is apt to cause confusion.
As a more appropriate way of expression, one may strongly advocate limitation
of the use of the word {\it phenomenon} to refer exclusively  to observations
obtained under specified circumstances, including an account of the 
whole experiment.}
\enq
Unfortunately, these  admonitions are  generally  ignored by the authors of 
{\it Quantum Enigma} (QE) \cite{bruce1}, 
and as a consequence this book 
will  cause plenty of the  confusion  predicted by Bohr.
In the following section, I will quote  selected paragraphs that contain what I consider to be
some of the more  outlandish statements which I found in this book,  followed by
my critique. In support of this critique, I include in
the last section some relevant quotations
from Einstein, Bohr, Heisenberg, and Schr\"{o}dinger, who laid the foundations
of modern quantum theory, and by some other prominent contemporary  physicists,
concerning  quantum mechanics and the measurement process.
These quotations contradict many of the  claims 
and unsupported assertions made in this book about the
interpretation  of quantum theory. In
particular, all these physicists  {\it deny} 
the supposed  role of  {\it consciousness}
in physics, for which there isn't any experimental
support whatsoever. But the claim that consciousness
plays a role in quantum mechanics (``our bias,'' as the authors write) is the underlying  message
which the authors of QE would like to implant on their  readers.

The authors of QE also offer quotations from prominent scientist
in  support for their claims. 
For example,  Martin Rees, the Astronomer Royal and current  president of the
Royal Society in London, is quoted as saying (ref. \cite{bruce1} page 193):

\beq
{\sl In the beginning there were only probabilities. The universe could only come into
existence if someone observed it. It does not matter that the observers turned
up several million years later. The universe exists because we are aware of
it.}\footnote{ The authors of QE do not
give sources for the numerous quotations in their book.
Therefore, I  asked Martin Rees where he had made
this statement and he responded:
 ``I am perplexed by the quote. It seems rather 'Wheelerish' -- not at all
the sort of thing I would have said.''  Apparently this  quotation was 
inserted by an editor as a caption to a figure  in an article that  
Rees had written for  the New Scientist (August 6, 1987). Indeed, it turns out
that this  figure is  a copy of a  drawing made by 
John Wheeler which also  appears on page 200 of QE, and that  the caption paraphrases 
Wheeler's  words in a script entitled ``Law without law."
The original caption starts with  ``The universe viewed as a self-excited circuit.'' 
Apparently, Wheeler was in a playful mood, 
because he remarked  ``caution: consciousness
has nothing whatsoever to do with quantum processes'' (see quotation at the
end of the next section)} (ref. \cite{bruce1} page 193).
\enq

In contrast, however,  Murray Gell-Mann, the winner of a Nobel Prize for
his fundamental contributions to particle physics, is quoted in QE as saying  that
\beq
{\sl The universe presumably couldn't
care less whether human beings evolved on some obscure planet to
study its history; it goes on obeying the quantum mechanical
laws of physics irrespective of observation by physicists} (ref. \cite{bruce1} p.156)
\enq

The authors' response is  that ``in talking
about classical physics, Gell-Mann's {\it presumption} [my emphasis]
would go without saying (ref. \cite{bruce1} p. 168).
But Gell-Mann is talking about quantum
mechanics which governs {\it all} the  fundamental  process  
in our the universe. The authors' distinction here
 between classical and quantum mechanics is a red herring. 
The  examples  discussed  in QE are concerned
with {\it elementary} or {\it single} quantum processes. Evidently, 
the authors are  unaware
that there is a distinction between analyzing  such processes 
and the multiple quantum processes that occur in macroscopic systems
like living creatures, stars, and galaxies.\footnote {For example, consider the evolution of a star
like the sun.  This evolution  is primarily
based on quantum phenomena, with classical gravitational forces primarily providing
the confinement for the atomic process.
Thus, the production  of energy
is due to nuclear reactions  governed  by quantum processes
in the interior of the star, which requires also  the phenomenom of  quantum tunneling,
the emission of radiation, which  is governed by the laws of quantum field theory,
and the eventual collapse of a star like the sun 
into a white dwarf or neutron star, which  depends on the
quantum mechanical  degeneracy pressure of electrons or neutrons \cite{weisskopf}.
Even the explosion of more massive stars into supernovas gives rise to 
quantum nuclear processes leading to the emission  of  neutrinos. 
Since the start of the Big Bang 
all these processes have been going on without the need of any ``observers,''
conscious or otherwise. The most striking evidence of the Big Bang is
the low-temperature black-body radiation which was created in the early universe
by quantum processes, obviously without  any  observers around \cite{michael}}

In fact, it is the statement made by  Gell-Mann and  not the one attributed
incorrectly to  Rees (see footnote 1),  
that represents the  generally accepted view that {\it all}  processes 
in the universe evolve 
in accordance with the laws of quantum mechanics {\it without} any need 
whatsoever of conscious observers.  In cases where classical mechanics 
is adequate to explain the observations, it is  regarded as 
an approximate theory.

John Bell, who is regarded as having made some 
of the most significant modern  contributions to our understanding of
quantum theory, remarked that

\beq
{\sl I see no evidence that it is so [that the cosmos depends on our being
here to observe the observables] in the success of contemporary
quantum theory. So I think {\rm it is not right} [my emphasis]
to tell the public that a central role for
conscious mind is integrated into modern atomic physics. Or that
`information' is the real stuff of physical theory \cite{bell2}

I think the experimental facts which are usually offered to show 
that we must bring the observer into quantum theory do not compel
us to adopt that conclusion. } \cite{bell1}
\enq

For  a long time  I have  argued along the same lines that 
I found recently in an article by  A. Leggett \cite{leggett}, 
a Nobel prize winner who has given
considerable thought to the quantum measurement paradox,
\beq
{\sl ...it may be somewhat dangerous to `explain' something one does not understand
very well [the quantum measurement process] by invoking something [consciousness]
one does not understand at all!}
\enq
But instead of ``it may be dangerous,'',  I would say ``it is nonsense.''

These  last three comments firmly  contradict  the central
claims of the authors of QE concerning  a supposed  role for consciousness 
in physics .\footnote { I share 
Steve Hawking's impulse
``to reach for my gun'' (ref. \cite{bruce1}, p. 120) 
whenever  Schr\H{o}dinger's cat story is told.  This cat story is
notorious. It requires  one to accept  that a cat, which can be
in innumerable  different states,  can be represented
by a two-state wavefunction, a bit of nonsense which Schr\H{o}dinger himself
originated. However, a movie camera installed in the box containing the cat
would {\it record} a cat that is alive  until the unpredictable moment that the 
radioactive nucleus decays opening the bottle containing cyanide  thus killing
the cat.   
It is claimed that Schr\H{o}dinger  never
accepted the statistical  significance of his celebrated wavefunction. 
}

\subsection* {Critique of selected quotations from QE }

Each of the  quotations from QE  listed below is  
written in typewriter style, and its location in QE  is identified by 
the page number (p) and the line number counted either from the bottom (b)
or the top (t) of the page. My criticisms  follow after  each 
of these quotations. 

\beq
{\tt This is a controversial book. But nothing we say about {\it quantum 
mechanics} is controversial. The experimental results we report and our
explanation of them with quantum theory are completely undisputed} 
(p.3 t.1)
\enq
Invariably, however,  experimental results  are  reported in this
book in a very  sketchy and innacurate manner,
which leads to the confusion  predicted by Bohr (see Introduction). Moreover, as will
be shown below, the statement that ``our explanation of them [the experiments]
with quantum theory is undisputed'' usually turns out to be false.

\beq
{\tt That physics has {\it encountered} consciousness cannot be denied}
(p.4  b.13) 
\enq
 Presumably Bohr, Heisenberg, Einstein, and many  other great
physicists  I quote  in this critique have been living in denial.
Only very  few prominent physicists, e.g.  E. Wigner,  have argued for a role
of consciousness in the measurement process. But such  exceptional
cases do not  justify ignoring  the warning of Bell
``that it is not right to tell the public that a central role for consciousness 
is integrated into modern physics''   
\beq
{\tt The quantum enigma, conventionally called the ``measurement problem,''
appears right up front in the simplest quantum experiment. (p.5 t.9)}
\enq
The measurement problem can be summarized as the question why macroscopic detectors
like a Geiger counter or a photographic plate, which are ultimately
made of atoms, are never found  in a {\it superposition} of quantum states.
The authors of QE claim that {\it consciousness} is what destroys this superposition.
But if  there are several observers,  whose consciousness is responsible for this
collapse of superposition? If both authors of QE look  at the {\it same time}
at a photographic film recording
an atomic event, who ``caused'' the  collapse of superposition?
And what happens when you have a thousand observers, as is often
the case in current high-energy physics when  often billions
of events are observed?

\beq
{\tt Try summarizing the implications
of quantum theory, and what you get sounds mystical... To account for the demonstrated
facts, quantum theory tells us that an observation of one object can instantaneously influence
the behaviour of another greatly distant object---{\it even if no physical force
connects the two}.
Einstein rejected such influences as ``spooky
interactions,'' but they have now been demonstrated to exist.} (p.12 b.14) 
\enq

The ``facts'' that have  been demonstrated are  {\it correlations} between 
distant particles which are predicted according to quantum mechanics
as  a  consequence of {\it conservation laws}.
For example, due to conservation of energy and
momentum in a two particle scattering event,  the observation of the  
momentum of one of the scattered particles  determines  the momentum 
of the other scattered particle if the initial state is
known.  This correlation is expected both  in classical
and in quantum theory, and there is nothing ``mystical'' about it. There are also 
correlations of polarizations
of entangled photons which are predicted by quantum theory that have been confirmed
by experimental observation.  These correlations are  consequences of the {\it conservation
law} of angular momentum.  The claim that Einstein {\it rejected} these   
correlations is false (for some quotations of his views of quantum theory see section II).

\beq
{\tt For example, according to quantum theory, an object can be
in two or many places at once---even far distant places. Its existence at the particular
place it happens to be found becomes an actuality  upon its (conscious) observation.}
(p.7  b.9)
\enq

Both statements are false.  Quantum theory is a theory that  predicts 
the probability of observing  
physical attributes of a particle, such as position and momentum.
The  probability of finding 
a particle in ``two places at once'' is {\it always} zero.  
The question can be asked  as to where a particle  is located in between
observations, but this question is {\it metaphysical}, and lies
outside the realm of scientific inquiry.
The claim  that it requires  {\it consciousness} to make the location
of an object an ``actuality,''  which is
repeated like a mantra throughtout QE, is not supported by any evidence,
and it is demonstrably false.

\beq
{\tt But they [our physics collegues] will find nothing 
{\it scientifically} wrong with what we say. The physics
facts we present are undisputed.}
(p.13 b.12)
\enq 
This claim, which is being promoted  in QE  to  an unsuspecting
public, is untrue, as will be demonstrated here in many examples.

\beq
{\tt The waviness\footnote{ This name was introduced by Bell,
and refers to the absolute square of the wavefunction} in a region is the
probability of {\it finding} the object in that region.
Be careful---the waviness is not the probability of the object being there.
There is a crucial difference! The object was not there until you found it there.
Your happening to find it there {\it caused} it to be there. 
This is tricky and the essence of the quantum enigma. 
}
(p.75 b.12)
\enq

The muddled second half of this  statement is incorrect. 
The absolute square of the wave function  gives the probability
for the outcome of an experimental observation such as the
localization of an atomic  particle.
A  particle  can be  localized
by an appropriate  recording device, a Geiger counter, a photographic plate, etc.,  independent
of any particular human observer. One need only to realize  that different
observers who examine such a 
recording all reach the same conclusion about the region of localization
to see that the remark ``your happening to find it there {\it caused}
it to be there'' is nonsense.
An  observer does  not {\it cause} the occurence of an
atomic event.  This is like believing that you can  bend spoons with your  mind.
No one has given evidence for such effects. 

The authors of QE claim that 
\beq
{\tt In quantum theory there is no atom in addition to the wavefunction of the atom.
This is so crucial that we say it again in other words. The atom's wave-functions and
the atom are the same thing; ``the wave function of
the atom'' is a synonym for ``the atom.''}
(p.77 t.6)

{\tt Since the wavefunction is synonymous with the atom itself, the atom
is simultaneously in both boxes. The point of that last paragraph
is hard to accept. That is why we keep repeating it. (p.103 t.3)}
\enq
   In other words, since  matter
consist of  atoms,  presumably while reading this nonsense
you are  sitting on a chair made up of wavefunctions (do you also feel
the vibrations?). 
Of course, the wavefunction of an atom is not a synonym (having the same meaning)
for the atom. The claim for such a
{\it synonym} is nonsense.  
Atoms have mass, charge, total spin, and  energy levels, 
which are {\it invariant} properties, while
the wavefunction describes the evolution of its dynamical variables.
 According to quantum theory, the  square of the
wave function gives  the probability that the  measurement process
yields  allowed values for a set of commuting  variables.
For this purpose it is necessary to study  an
ensemble of atoms which initially are  prepared
under identically the same physical conditions.
This is fundamentally  different from claiming that the  wavefunction is {\it synonymous} 
with the atom itself. The probability of observing
an  atom {\it simultaneously} at two different locations, by an actual  measurement,
is always {\it zero}. Hence, it is false to claim
that the atom ``is {\it simultaneously} in both boxes.''

\beq
{\tt Accordingly, before a look collapses a widely spread-out  wavefunction
to the particular place where the atom is found, the atom did not exist there prior to the
look. The look brought about the atom's existence at that particular
place---for everyone.} (p.77 t.9)
\enq 
This is a  powerful ``look'' indeed, but again, this is false.  
In spite of Bohr's repeated  warning of the importance
of carefully choosing appropriate language to describe the
observation of atomic events,  this warning is again ignored here. 
Statements like ``before a look collapses ... a wavefunction'' or  
``the look brought about...the atom's existence'' do not make any
sense or have any meaning whatsoever. 
According to the description  of some of  the founders of the quantum theory  
who are quoted in the next section, a meaningful  statement is that 
the  reduction or collapse of a wavefunction 
occurs after a recording has been made by  an {\it irreversible}
amplification of an atomic event by  a  macroscopic detector,  like a Geiger counter
or a photographic plate. The combination of eye lens, retina, optical nerve and neural
memory cells can  be regarded as a detector  for the special case of
photons in a visible frequency range, but  such a detector is unique to a single observer.

\beq
{\tt The most accurate way of describing the state of the unobserved atom is to put
into English the mathematics describing the state of the atom before we looked
to see where it is: The atom was {\it simultaneously in two states}: in the first state,
it is in-the-top-box-and-not-in-the-bottom-box, and simultaneously
in the second state it is in-the-bottom-box-and-not-in-the-top-box.}
(p.79 t.15)
\enq
This so called ``accurate'' description is actually pure metaphysics, since it
is not possible to established the location of an ``unobserved'' atom.
According to  quantum theory, before an observation
is made  there is only  a {\it probability} of finding the atom in either the top
or the bottom box. The essence of quantum physics is that nothing definite 
can be said
about the location  of the atom before a measurement of it
is made. For example, we can send a beam of  photons  into a box  and detect
the  scattered photons. This can establish
approximately the localization of the atom {\it at the time} of the interaction
of the photon and the atom, but not at any time before this interaction
took place.
\beq
{\tt The talk we offer in a quantum mechanics class for physics
students is that when we look in a box and find no atom, we
instantaneously collapse the atom's waviness [wavefunction] into
the other box.} (p.108  t.11) 
\enq
The statement refers to the possibility 
that even though a detector is not triggered  
 the wavefunction of the system is changed.
This appears to contradict the requirement
that the detector must make a recording
to alter the  wavefunction of the system.
In fact, the presence of a detector to determine, for
example, the location of an atom, also  alters
the evolution of the wavefunction, and consequently
the probabilities of various outcomes. Therefore,  
if a detector is not triggered, there is a finite
probability that the atom  did not take the path
crossing the detector. But there is also a non-vanishing
probability that the atom crossed the detector 
without triggering it. Therefore, it can not be
concluded with  certainty in what box the atom is
located until a second   detector triggers and  determines
its location.  Certainly  {\it consciousness}, i.e. `` we look
in a box and find no atom, we instanteniously
collapse the atom's waviness ...'' , has nothing whatsoever 
to do with the explanation of this subtle property of quantum mechanics.

\beq
{\tt But randomness was not Einstein's most serious problem with quantum mechanics.
What disturbed Einstein , and more people today, is quantum
mechanics' apparent denial of ordinary physical reality---or, maybe the same
thing, the need to include the observer in the physical description---an
intrusion of consciousness into the physical world.}
(p.80 t.8)
\enq
While  Einstein was concerned with the 
nature of physical reality, it is highly misleading to indicate that it is
the ``same thing'' to imply 
that he was concerned with the ``intrusion of consciousness into the physical
world.'' On the contrary, he made it clear that the generally
accepted statistical interpretation of the wavefunction
is an ``{\it objective} description'' whose concepts clearly
make sense independently of the observation and the {\it observer} (see the next section
for the full quotation in his letter to Born a year before his death).  

\beq
{\tt John Bell  felt that the quantum mechanical description will be
superseded... It carries in itself the seeds of its own destruction... He feels
that ``the new way of seeing things will involve an imaginative leap that will
astonish us.''}
(p.87 t.5)
\enq
The authors of QE fail to point out that this quotation appeared  
in an article written jointly by Bell and myself about
40 years ago \cite {michael1}. But neither of them  bothered to
ask for my current  views on this subject. I will take thi
opportunity  
to make a disclaimer here.  In our paper we also said, tongue in cheek, that
\beq
{\sl The experiment may be said to start with the printed proposal and
to end with the issue of the report. The laboratory, the experimenter,
the administration, and the editorial staff of the {\rm Physical Review}
are all just part of the instrumentation. The incorporation of (presumably)
conscious experimenters and editors into the equipment raises a very intriguing
question... If the interference is destroyed,  then the Schr\H{o}dinger equation
equation is incorrect for systems containing consciousness. If the interference
is not destroyed, the quantum mechanical description is revealed as not wrong
but certainly incomplete.}
\enq
In a footnote we added that 
 ``we emphasize not only that our view  
is that of a minority
but also that current interest in such questions is small. The typical
physicist feels that they have been long answered, 
and that he will fully understand
just how if ever he can spare twenty minutes to think about it.''
(As a graduate student, one of the authors, Bruce Rosenblum, reports that 
he ``assumed  that if  I spent an hour or so thinking it 
[the wave particle duality] through, I'd see it all clearly'' ( ref. 
\cite{bruce1}, p. 10).)
                                                                                             
At the time that we made these comments,
I did not understand   that a Geiger counter, a photographic film,
the retina of the eye and associated neural connections.
or any other detector creates a more or less permanent  record
by means of  physical and/or  chemical processes that are {\it  irreversible}.
As in thermodynamics, such  processes involve a very large number of atoms
which are involved through an amplification process  essential to 
the creation of a recording.\footnote {Following Bohr's admonition to
discuss the details of the measuring apparatus,
we note  that a typical photographic film consist of an irregular
array of silver bromide ionic crystals suspended on an emulsion. The absorption of a photon
ejects an electron which gives rise to a catalytic  reaction where a very large number
of silver ions transform into silver atoms. After the film is processed
these  silver atoms give rise to a  spot which absorbs visible light
shining on it, thus
producing  a dark spot  which approximately marks  the
position on the screen where the photon had landed. The cascade of silver
ions into silver atoms produced by the  photon is the amplification
which which gives rise to an
irreversible process, and therefore it can not be described 
by the evolution of a simple  wavefunction.
}

A major failure of QE  is that 
this essential feature of the measurement process in
quantum theory  is not even mentioned.
It is worth remembering 
 that irreversibility  already was a conundrum for classical physics
because the basic equations are time reversal invariant. The same
invariance applies to the basic equations in quantum mechanics. The emergence
of irreversibility occurs in systems with large number of particles,
which is the case  with  recording devices, 
and constitutes the foundations of statistical thermodynamics \cite{michael}.

\beq
{\tt There is no official Copenhagen interpretation.  But every version grabs the bull
by the horns and asserts that an {\it observation produces the property observed}.}
(p.100 t.11)
\enq

Read, for example,  the quotations of Bohr, Heisenberg, Einstein and others quoted 
here and in the next section to see that none of them ever made such a  nonsensical statement  as 
``produces the property observed'' when referring to an observation. 

\beq
{\tt Even students completing a course in quantum mechanics, when asked what the
wavefunction tells, often incorrectly respond that it gives the probability where
the object is.}
(p.103 t.7)
\enq
Actually,  the students'   response is essentially correct if it is 
somewhat modified  by stating  that 
 ``it is the probability  where the object will be  found '' 
in an experiment designed
to establish the position of the object.  

\beq
{\tt Our concern is with the consciousness central to the quantum enigma---the awareness
that appears to affect physical phenomena. Our simple example was that your
observation of an object wholly in a single box {\it caused} it to be there, 
because you presumably {\it could} have chosen to {\it cause} an interference
pattern establishing a contradictory situation, whereby the object would have
been a wave simultaneously in  two boxes.} 
(p.168 t.18)
\enq
It is hard to imagine that more nonsense about the meaning of 
quantum theory could  be encapsulated into a single sentence.
To begin with, an experimental set up to  observe in which of two boxes  an atom is
located does not ``cause'' such a  localization. Instead, an observer
who examines the output of a recording  finds the
location of the atom in accordance with a probability distribution given by the absolute
square of Schr\"{o}dinger's  $\psi$ function 
appropriate to the experimental arrangement.
A different experimental setup may lead to  observations of interference
patterns by coherent atomic beams, but these patterns 
are not ``caused'' by the ``awareness'' of the
observer. The interference patterns can be recorded on a photographic film
and  seen there  by anyone examining this film. 
It is completely ridiculous to propose that someone  who sees
the interference pattern on the film ``caused'' it to be there.
The  claim that by ``choosing'' the experimental
set up the observer ``caused'' a particular outcome is a bizarre
use of the word ``cause'' which has nothing whatsoever to do with
quantum theory. Bohr must be turning over in his grave.

According to the authors of QE,
{\it consciousness} is supposed to enter into quantum theory  because the observer
can  choose (``free will'') the experimental setup, which  demonstrates in this case
complementary aspects,  wave or particle behaviour,  of a photon, electron
or other atomic objects. But the choice of experiment can also be made, for example, by 
flipping a coin.

\beq
{\tt Does such a demonstration necessarily require a {\it conscious} observer?
Couldn't a conscious  robot, or even a Geiger counter, do the observing?
That most commonly voiced objection to consciousness being required comes up---and
is refuted---in our next chapter. For now, just recall that, according
to quantum theory, if that robot or Geiger counter were not in contact with
the rest of the world it would merely entangle to become part of a total superposition
state---as did Schr\"{o}dinger's cat. In that sense it would not truly {\it observe}}
(p.168 b.12)
\enq
The operation of a Geiger counter is well understood, and it satisfies the
condition necessary  to establish an  observation, namely
the irreversible  amplification of an atomic
signal to create a semi-permanent  record of an atomic event. 
The only important contact  of a Geiger counter to the ``rest of the  world'' 
is a power cable, which has to be plugged into a wall socket from which
electricity flows from a power plant.  A battery would also 
be sufficient. 

\beq
{\tt Classical physics, Newtonian physics, is completely deterministic. 
An ``all-seeing eye,''
knowing the situation of the universe at one time, can know its entire future. 
If classical physics applied to everything, there would be no place for free will.}
(p.169 t.11)
\enq

This familiar  quotation, due to Pierre Simon Laplace in the 18{\it th}
century,  was  later shown 
to be false by Henri Poincar\'{e}. Laplace was not aware that Newtonian
mechanics  could lead to chaotic motion, and that in practice
classical theory also is  not
deterministic,  because of ``sensitivity to initial conditions.''\footnote{ Doubters 
should play with the double pendulum in the entrance
of our physics dept., which I had built by our mechanics shop to demonstrate 
chaotic dynamics.}. The implication that quantum mechanics is somehow essential
to {\it free will}, which is supported by some philosophers of science, is not
valid.\footnote{Some time ago I had an argument about this issue with 
the Berkeley philosopher John Searle.
However, he did not seem to be aware that 
the {\it noise} due to finite temperature
thermal fluctuations
in the firing of neural brain cells is more important than the  
zero-temperature quantum fluctuations or effects of the uncertainty principle.}
\beq
{\bf The Encounter ``Officially'' Proclaimed (p.180 t.1)}
\enq

\beq
{\tt In his rigorous 1932 treatment ``The Mathematical Foundations of Quantum
Mechanics,'' John von Neumann showed that quantum theory makes physics encounter
with consciousness {\it inevitable} [my emphasis]} (p.180 t.1)
\enq
But in the next sentence,  one  finds that this ``inevitable encounter'' occurs
because von Neumann has treated a Geiger counter by a trivial  wavefunction 
consisting of the superposition of only two states: 
whether it  is in a  ``fired'' or in an ``unfired'' state.
This model of a Geiger counter, however,  is {\it incorrect}, because it does not
describe  the {\it essential} property of such a  detector, which is to  
be able to make a permanent  {\it record} of
an atomic event. Such a  recording requires  an {\it irreversible} process. 

\beq

{\tt You're prompted to investigate how the robot chose which experiment
to do in each case. Suppose that you find that it flipped a coin. Heads
it did the-look-in-the-box experiment; tails the interference experiment.
You find something puzzling about this: The coin's landing seems inexplicably
connected with what was  presumably in a particular box-pair set.
Unless ours ia a strangely deterministic world, one that conspired to correlate
the coin landing with what was in the box pairs, there is no physical mechanism
for that correlation.}
(p.183 b.15)
\enq
The  outlandish notion that there exists a
correlation between the  ``coin landing'' and the outcome of an experiment is
tied to the previous bizarre claim that an observer can ``cause''  this 
outcome. This is the consequence of the sloppy  language which Bohr had 
warned should be avoided.

\beq
{\tt You therefore replace the robot's coin flipping by the one decision mechanism
you are sure is not connected with what supposedly exists in a particular box-pair
set: {\it your own free choice}. You push a botton telling the robot which
experiment to do with each box-pair set. You now find that by your
conscious free choice of experiment you can prove either that the objects
were concentrated or that they were distributed. You can choose to prove either
of two contradictory  things. You are faced with the quantum enigma, and
consciousnes is involved.}
(p.183 b.8)
\enq
If you have had the patience to come this
far,  you finally will have found out how, according to
the authors of QE,  {\it physics encounters consciousness}: by 
replacing the  decision making of a robot  who  flips  coins
with the {\it free will of a conscious observer}. This claim, however, does not make any
sense, because as a ``conscious observer'' you can also make decisions based on
flipping coins rather than on your free will. Following  the logic
of the authors of QE, you would  have to conclude
that {\it physics encounters coin flipping}. It should be pointed out that
while a few eminent physicists such as Eugene Wigner believe that consciousness 
plays a role in the 
{\it collapse} of the wavefunction, he did  not argue that this role 
has anything to do with ``free will''. This is a new and unsubstantiated
twist to the role of consciousnes given by the authors of QE.
Such a  bias should not be foisted on the general public
as a generally accepted notion in modern physics.\footnote{
It is easy to demonstrate that the persistent claim  of the
authors of QE that it is necessary to choose or
flip a coin to decide which of their two  
{\it gedanken} experiments should be  performed
can be avoided by performing {\it both} experiments at the
same time. Light  of appropriate frequency and variable intensity is shined  on  the 
two slits through which individual  atoms of fixed momentum are directed. 
Then  a  scattered photon and an appropriate detector can  
record through which slit an atom passed, while   a photographic plate can record
the position where  the atom lands later on. All the spots  that are recorded 
on the photographic plate are
then separated into two classes of events: a) events where a scattered photon is detected
and b) events where no scattered photon is detected. According to  quantum theory,
 events of class b) show an  interference pattern characteristic of the two
slit experiment although it is modified by the presence of the
radiation field.  Events of class a), however,  do not show
this interference pattern. The fraction of the events in class a) and b) depends
on the intensity of the light shining on the two slits: the weaker the intensity 
the greater the number of events in class a) and viceversa.
It is assumed here that the detector of  scattered photons  is very
efficient, otherwise the interference pattern will be further smeared out,
because it contains also events where a scattered photon from an atom was not detected.
See also the description given by R. Feynman in reference \cite{feynman2} \cite{feynman}}

\beq
{\tt If we consider the robot argument from the viewpoint of quantum theory, 
the isolated robot is a quantum system, and  von Neumann's conclusion applies:
the robot entangles with the object in the box pairs, and the object's wavefunction
does not collapse into a single box until a conscious observer views the robot's
printout.}
(p.184 t.1)
\enq
The notion that the the ``robot entangles with the object,''
implying that
the robot, or for that matter any macroscopic measuring apparatus like a Geiger counter
or a photographic film, can be represented by a wavefunction $\psi$ is
{\it incorrect}. The essential property of  macroscopic  systems 
like robots is that these systems
are able to create a record of an atomic event by a mechanism of amplification
which is irreversible. This mechanism
has nothing whatsoever to do with consciousness.

In conclusion, I like to  quote John Wheeler \cite {wheeler}  
\beq
{\sl Caution: ``Consciousness'' has nothing whatsover to do with the quantum process.
We are dealing with an event that makes itself known by an irreversible
act of amplification, by an indelible record, an act of registration.
Does that record subsequently enter into the ``consciousness'' of some person,
some animal or some computer? Is that the first step into translating
the measurement into ``meaning''---meaning regarded as ``the joint product of
all the evidence that is available to those who communicate.'' Then that is
a separate part of the story, important but not to be confused with
``quantum phenomena.''}
\enq

This quotation  appears in QE, p. 165,  
but in a truncated form.  
What is left out is  Wheeler's 
definition for the word {\it meaning}.
The authors of QE comment to Wheeler's remark 
\beq
{\tt We take this as an injunction to physicists (as physicists) to study only
the quantum phenomena, not the {\it meaning} of the phenomena.}
\enq
But Wheeler's {\it injunction} 
clearly is {\it against}  claiming that {consciousness} 
has something to do with 
quantum phenomena, which  is the central theme of QE. It is not an injunction
against studying the cognitive processes in the brain associated with
consciousness, which are
``a separate part of the story, important but not to be confused  with
quantum phenomena.'' 

\subsection*{Remarks  on the quantum  measurement process 
 by some of the founders of the quantum theory,
 and by prominent contemporary  physicists}

Below, I have included some quotations on quantum measurement and the role of
the observer  by the founders of quantum theory
and some prominent contemporary  physicists, which are relevant
to my  critique of QE.\\ 

N. Bohr (1958)
\beq
{\sl Far from involving any special intricacy, the irreversible amplification effects which
the recording of the presence of atomic objects rests rather reminds of the essential
irreversibility  inherent in the very concept of observation. The description of atomic
phenomena has in these respects a perfectly {\rm objective} [my emphasis] character, 
in the sense that no
explicit reference is made to any individual observer and that therefore, with proper
regard to relativistic exigences, no ambiguity is involved in the communication of
information} \cite {bohr1}
\enq

W. Heisenberg (1958)
\beq
{\sl The probability function does---unlike the common procedure in
Newtonian mechanics---not describe a certain event but, at
least during the process of observation, a whole {\rm ensemble}
[my emphasis] of events...When the old adage ``{\rm Natura non facit saltus}''\footnote{
Nature does not act in jumps.}  
is used as a basis of criticism of quantum theory, we can reply that
certainly our knowledge can change suddendly and that this fact justifies
the use of the term ``quantum jumps'' [or collapse of the wavefunction].

Therefore, the transition from the ``possible'' to the ``actual''
takes place during the act of observation. If we want to describe
what happens in an atomic event, we have to realize that the word
''happens'' can apply only  to the observation,
not to the state of affairs between two observations. It applies
to the physical, not the {\rm psychical} [my emphasis] act of  observation,
and we may say that the transition from the ``possible'' to the ``actual''
takes place as soon as the interaction of the object with the measuring
device, and therefore with the rest of the world, has come into play; it
is not connected with the act of registration of the result by the
{\rm mind} [my emphasis] of the observer. The discontinuous change
in the probability function [collapse of the wavefunction], however,
takes place with the act of registration, because it is the discontinuous
change of our knowledge in the instant of registration that has its
image in the discontinuous change of the probability function.
...Certainly quantum theory  does not contain genuine subjective
features, it does not introduce the {\rm mind} [my emphasis] of the
physicist as a part of the atomic event}  \cite{heisenberg1}.
\enq

A. Einstein (1949)

\beq
{\sl One arrives at very implausible theoretical conceptions, if
one attempts to maintain the thesis that the statistical quantum
theory is in principle capable of producing a complete description
of an {\rm individual} [my emphasis] physical system.
On the other hand, those difficulties  of theoretical interpretation
disappear, if one views the quantum mechanical description
as the description of ensembles of systems.

I reached this conclusion as the result of quite different
types of considerations. I am convinced that everyone who
will take the trouble to carry through such reflections
{\rm conscientiously} [my emphasis] will find himself finally
driven to this interpretation  of quantum-theoretical
description (the $\it \psi$ function is to be understood
as the description not of a single system but of an ensemble
of systems)} \cite{einstein1}
\enq

A. Einstein (1953)
\beq
{\sl All the same it is not difficult to regard the step into probabilistic
quantum theory as final. One only has to assume that the $\psi$ function relates
to an {\rm ensemble} [my emphasis], and not to an individual case...
The interpretation of the $\psi$ function as relating to an  ensemble
also  eliminates the paradox that a measurement carried out in {\it one} part of space
determines the {\it kind} of expectation for a measurement carried out later in
{\it another} part of space (coupling of parts of systems far apart in space)} \cite{einstein3}
\enq

A. Einstein (1954)
\beq
{\sl The concept that the $\psi$ function {\rm completely} describes
the physical behaviour of the individual single system is untenable}
\footnote{ Einstein reached this conclusion by the following argument:
\beq
My assertion is this: the $\psi$ function cannot be regarded as a complete
description of the system, only as an incomplete one. In other words:
there are attributes of the individual system whose reality no on doubts
but which the description by means of the $\psi$ function does not include.

I have tried to demonstrate this with a system which contains one `macro-coordinate'
(coordinate of the centre of a sphere of 1 mm diameter). The $\psi$ function selected
was that of fixed energy. This choice is permissible, because our question by its
very nature must be answered so that the answer can claim validity for every $\psi$ function. 
From the considerations of this simplest case, it follows that---apart from
the existing macro-structure according to the quantum theory---at any arbitrarily chosen
time, the centre of the sphere is just as likely to be in one position (possible in
accordance with the problem) as in any other. This means that the description by
$\psi$ function does not contain anything which corresponds with a (quasi-) localization
of the sphere at the selected time. The same applies to all systems where macro-coordinates
can be distinguished.

In order to be able to draw a conclusion from this as to the physical interpretation
of the $\psi$ function we can use a concept which can claim to be valid independently
of the quantum theory and which is unlikely to be rejected by anyone: any system is
at any time (quasi-) sharp in relation to its macro-coordinates. If this were not the
case, an approximate description of the world in macro-coordinates would obviously
be impossible ('localization theorem'). I now make the following assertion: if the description
by a $\psi$ function could be regarded as the complete description of the physical condition
of an individual system, one should be able to deduce the `localization theorem' from
the $\psi$ function and indeed from any $\psi$ function belonging to a system which
has macro-coordinates. It is obvious that this is not so for the specific
example which has been under consideration \cite{ einstein2}
\enq
Stated in another way, the Schr\H{o}dinger equation has solutions for $\psi$, such as
a plane wave (fixed energy), which do not correspond to the description
of the motion of any individual macroscopic object like Einstein's  sphere
which has also sharply localized position. But this solution does describe
an {\it ensemble} of identical spheres  which have centers  
distributed uniformly along the direction
of motion, all moving with the same velocity.}
But one can well make the following claim: if one regards the $\psi$
function as the description of an {\it ensemble} it furnishes statements
which---as far as we can judge---correspond satisfactorily to those of classical
mechanics and at the same time account for the quantum structure of reality.
In this interpretation [Born's statistical interpretation of quantum
mechanics] the paradox of the apparent coupling of spatially separated
parts of systems also disappears. Furthermore, it has the advantage that
the description thus interpreted is an {\it objective} description whose
concepts clearly make sense independently of the observation and
the {\rm observer} [my emphasis] \cite{einstein2}
\enq

R.P.  Feynman (1963)
\beq
{\sl Nature does not know what you are looking at, and she behaves
the way she is going to behave whether you bother to take down
the data or not} \cite{feynman2}
\enq

J. Wheeler (1986)
\beq
{\sl No elementary quantum phenomenon  is a phenomenon until it's brought
to a close by an irreversible act of amplification by a detection
such as the click of a geiger counter or the blackening of a a
grain of photographic emulsion}\cite{wheeler1}.
\enq

J. Bell (1986)
\beq
{\sl I think the experimental facts which are usually offered to show 
that we must bring the observer into quantum theory do not compel
us to adopt that conclusion. \cite{bell1}

The problem of measurement and the observer is the problem where 
the measurement begins and ends, and where the observer begins
and ends...there are problems like this all the way from the retina
through the optic nerve to the brain and so on. I think, that---when
you analyse this language that the physicists have fallen
into, that physics is about the result of observation---you find
that on analysis it evaporates, and nothing very clear is being said.}
\cite{bell1}
\enq

N. van Kampen (1988)
\beq
{\sl Theorem IV. Whoever endows $\psi$ with more meaning than is needed
for computing observable phenomena is responsible for the consequences...}
\cite{kampen}
\enq


\begin{thebibliography}{30}
                                                                                                     
\bibitem{bohr2}              Quoted in A. Pais {\it Subtle is the Lord...}
                             (Clarendon Press, New York 1982) p. 455

\bibitem{bruce1}             B. Rosenblum and F. Kuttner, {\it Quantum Enigma, Physics
                             Encounters Consciousness} (Oxford Univ. Press, 2006)

\bibitem {weisskopf}         M. Nauenberg and V.F. Weisskopf, ``Why does the Sun shine,''
                             Am.. J. Phys. {\bf 46}, 23 (1978)

\bibitem {michael}           M. Nauenberg, ``The evolution of radiation towards
                             thermal equilibrium: A soluble model which illustrates
                             the foundations of statistical mechanics,''
                             Am. J. Physics {\bf 72}, 313 (2004)

\bibitem{bell2}              J.S. Bell, ``Speakable and unspeakable in quantum
                             mechanics,'' Introductory remarks at Naples--Amalfi meeting,
                             May 7, 1984.  
                             Reprinted in  J.S. Bell, {\it Speakable and
                             Unspeakable in Quantum Mechanics} (Cambridge Univ. Press
                             1987) 

\bibitem{bell1}             P.C.W. Davies and J.R.Brown, 
                            {\it  Ghost in the Atom} (Cambridge Univ. Press.
                             1986)  Interview with J. Bell, pp. 47-48

\bibitem{leggett}            A. Leggett, ``Reflections on the quantum measurement
                             paradox,'' in {\it Quantum Implications}, edited by ???
                             (Routledge, London 1991) p.94
 

\bibitem{bohr1}             N. Bohr, ``Quantum Physics and Philosophy, Causality and 
                                       Complementarity''
                            in {\it Essays 1958--1962 on Atomic Physics and Human Knowledge}
                            (Vintage Books, 1966)

\bibitem{heisenberg1}       W. Heisenberg, {\it Physics and Philosophy, The Revolution in
                            Modern Science} (Prometheus Books, New York, 1999) p. 55

\bibitem{einstein1}         {\it Albert Einstein: Philosopher--Scientist}\\
                            v.II, edited by P.S. Schilpp (Harper, New York 1949)
                            p. 67

\bibitem{einstein3}         A. Einstein in a letter to Max Born on Dec. 3, 1953.
                            Reprinted in {\it The Born--Einstein letters
                            1916--1955},  translated into English by Irene Born
                            (Macmillan Press, China 2005) p. 204

\bibitem{einstein2}         A. Einstein,  Unpublished commentary sent in a letter to Max Born
                            on 12 January 1954. Reprinted in {\it The Born--Einstein Letters
                            1916--1955},  translated into English by Irene Born
                            (Macmillan Press, China 2005) pp. 210-211

\bibitem{wheeler1}         P.C.W. Davies and J.R.Brown, interview with J.A. Wheeler, p.61, in
                            {\it  Ghost in the Atom} (Cambridge Univ. Press.
                             1986) 

\bibitem{wheeler}            J. Wheeler, ``Law without law'' in {\it Quantum
                             Theory and Measurement}, edited by J.A. Wheeler and
                               Zurek (Princeton Univ. Press, 1983) p. 196


\bibitem{michael1}           J.S.  Bell and M. Nauenberg, ``The moral aspects
                             of quantum mechanics,'' in {\it Preludes in Theoretical
                             Physics}, edited by A. De Shalit, Herman Feschbach,
                             and Leon van Hove (North Holland, Amsterdam 1966), pp.
                             279-286. Reprinted in  J.S. Bell {\it Speakable and
                             Unspeakable in Quantum Mechanics} (Cambridge Univ. Press
                             1987) p. 22 

\bibitem {feynman2}         R.P. Feynman, R.B. Leighton and M Sands, 
                           {\it The Feynman lectures on Physics}\\
                            vol. 3 (Addison Wesley, Reading 1965) 3-7

\bibitem {feynman}           R.P. Feynman, {\it Theory of Fundamental Processes}:
                             Notes  of Feynman's 1956 lectures at
                             Cornell University  taken by Peter Carruthers and Michael Nauenberg
                             (Benjamin, New York 1961), pp 1-2 


\bibitem{kampen}           N.G. van Kampen, `` Ten theorems about quantum mechanical measurements''
                           Physica {\bf A 153} 97 (1988)

\end{thebibliography}
\end{document}